\documentclass{aa}

\usepackage{graphicx}
\usepackage{float}
\usepackage[varg]{txfonts}
\usepackage{amsmath}
\usepackage{hyperref}

\newcommand{\msunyr}{\ensuremath{M_{\odot}\,\mathrm{yr}^{-1}}}
\newcommand{\kms}{\ensuremath{\mathrm{km\,s}^{-1}}}

\newcommand{\pcg}{PCG\,11}

\usepackage{cuted}   % per lo strip a piena larghezza
\usepackage{capt-of} % per didascalie fuori da un float

\begin{document}

\title{A Wolf-Rayet nebula in its disruption phase:\\ high-resolution radio observations of PCG 11}
\titlerunning{The Wolf-Rayet nebula PCG 11}
\authorrunning{Buemi et al.}

\author{C.S.~Buemi\inst{1},
%        \and Co-authors
P.~Leto~\inst{1}, 
F.~Cavallaro~\inst{1},
G.~Umana~\inst{1}, 
C.~Bordiu~\inst{2,1},  
F.~Bufano~\inst{1},  
C.~Trigilio~\inst{1}, 
M.D.~Filipovi\'c~\inst{3},
L.~Cerrigone~\inst{4,5},   
R.Z.E.~Alsaberi~\inst{3,6},
A.~Ingallinera~\inst{1},      
S.~Loru~\inst{1}, 
S.~Riggi~\inst{1}, 
A.C.~Bradley~\inst{3},
Z.J.~Smeaton~\inst{3}
        }

\institute{INAF -- Osservatorio Astrofisico di Catania, Via S. Sofia 78,
I-95123 Catania, Italy\\
\email{carla.buemi@inaf.it} %% CHECK email
\and
%Bordiu
Instituto de Astrof\'{i}sica de Andaluc\'{i}a (IAA-CSIC), Glorieta de la Astronom\'{i}a s/n, 18008, Granada, Spain
\and
%Filipovic - Bradley - Smeaton
Western Sydney University, Locked Bag 1797, Penrith South DC, NSW 2751, Australia
\and
%Cerrigone
National Radio Astronomy Observatory, 520 Edgemont Road, Charlottesville, VA 22903, USA
\and
%Cerrigone
Joint ALMA Observatory, Alonso de C\'{o}rdova 3107, Vitacura, Santiago, 7630355 Chile
\and
%Alsaberi Rami
Faculty of Engineering, Gifu University, 1-1 Yanagido, Gifu, 501-1193, Japan
%[Other institutes]
}

%\date{Received [XX]; accepted [XX]}

\abstract
%%%%%%%%%%%%%%%%%%%%%%
% context
{
Wolf-Rayet stars have high mass-loss rates and fast stellar winds. 
Their interaction with circumstellar material from previous evolutionary phases can produce complex nebular morphologies. 
Hydrodynamic instabilities can fragment wind-swept material, producing condensations, filaments, and gaps that alter the confinement of stellar wind and ionizing radiation. 
The role of fragmentation in the late evolution and subsequent dispersal of WR nebulae remains poorly constrained. 
PCG~11, with its fragmented morphology and regular scalloped rim, offers an opportunity to investigate these processes.
}
% aims
{
We aim to characterize the structure of PCG~11 and investigate its ring-like morphology and scalloped rim. 
We constrain its three-dimensional geometry and determine whether the optical fragmentation pattern is also traced by ionized gas.
}
%%%%%%%%%%%%%%%%%%%%%%
% methods
{
We obtained high-resolution radio continuum images with MeerKAT at 2.6 and 3.1 GHz and ATCA at 5.5 and 9.0 GHz. 
We used radio morphology to constrain nebular geometry and analysed the azimuthal structure of its rim.
}
%%%%%%%%%%%%%%%%%%%%%%
% results
{
The radio emission reveals a hollow, equatorially enhanced structure viewed close to its symmetry axis. 
The images resolve the rim into a coherent sequence of condensations, low-brightness gaps, and finger-like protrusions. 
We quantify the fragmentation scale and relate it to the independently constrained nebular geometry. 
We also detect an unresolved source coincident with WR~75ab, whose rising radio spectrum is consistent with thermal emission from the stellar wind. 
The nebula processes only a small fraction of the stellar ionizing radiation, indicating substantial photon leakage.
}
%%%%%%%%%%%%%%%%%%%%%%
%%%%%%%%%%%%%%%%%%%%%%
% conclusions
{
\object{PCG 11} appears to capture a poorly observed stage in the evolution of a massive-star bubble: 
the swept-up shell has developed a coherent instability pattern and become fragmented and porous, while remaining dynamically coupled to the stellar wind. 
It provides a clear view of how fragmentation and erosion may reshape the circumstellar environment during a transitional phase in the evolution of wind-blown nebulae around massive stars.
}
%%%%%%%%%%%%%%%%%%%%%%


\maketitle
%\nolinenumbers

%-------------------------------------------------------------------
\section{Introduction}
Massive stars profoundly shape their environment through strong
stellar winds and ionizing radiation. Wolf-Rayet (WR) stars represent a particularly powerful wind phase in their evolution, with wind velocities of $v_\infty \sim 1000$--3000~\kms and high mass-loss rates \citep{Crowther_07}. 
Such fast winds can sweep 
up the slower, denser material shed during the previous evolutionary stages,
carving  shells and bubbles \citep[e.g.][]{chu_81,Tola_11}, whose morphology retains information on the previous mass-loss history and also influence how the mechanical energy of the current wind is transferred to the surrounding medium.
During the evolution of the swept up shell, hydrodynamical instabilities can fragment it, changing the confinement of the shocked wind and the coupling of mechanical energy to the surrounding gas \citep[e.g.][]{Garcia-Segura_96,dwarkadas22,dwarkadas23}.  Numerical models show that the fraction of the stellar mechanical energy retained by the circumstellar gas can be substantially lower than predicted by simple energy-conserving bubble models \citep{freyer06,dwarkadas23}.

When and how a swept-up shell fragments is therefore an important aspect of understanding how massive-star winds transfer energy and momentum to their surroundings. 
Although the formation of wind-blown bubbles around massive stars has been extensively investigated both theoretically and observationally 
\citep[e.g.][]{weaver77,toala15}, 
 much less is known about shell fragmentation and dispersion, that is a short-lived process difficult to identify observationally.

How efficiently stellar winds transfer mechanical energy to the surrounding gas is also relevant on the larger scales of young stellar clusters and associations, where the combined winds of massive stars constitute an important source of mechanical feedback during the early evolution of the cluster. The powerful winds of WR stars can make a significant contribution during this phase, although in crowded environments the effects of individual stars are difficult to disentangle,
Westerlund~1 is an example of combined effect of WR winds (Umana et al. in prep).
More isolated WR nebulae provide a simpler view of the wind-shell interaction. Radio continuum observations are particularly valuable in this respect, as they trace the ionized gas independently of dust extinction and, at high angular resolution, resolve the structure of the swept-up shell. They can also detect the thermal emission from the stellar wind itself, thus probing both the current mass loss and the material with which the wind interacts.
The combination of sensitivity, angular resolution, and broad spatial-scale coverage offered by modern radio interferometers 
is revolutionizing the study of evolved stars, allowing for fast snapshots that reveal numerous circumstellar structures at once (Umana et al. in prep, Buemi et al. in prep, and \citealp{Bordiu2025}).
MeerKAT currently provides these capabilities up to S-band frequencies, while higher-frequency observations can be obtained with facilities such as the Australia Telescope Compact Array (ATCA). The advent of SKA-Mid will bring  high-sensitivity, broad spatial-scale imaging capabilities to frequencies up to 15.4 GHz, opening new possibilities for spatially resolved studies of the interaction between massive-star winds and their circumstellar nebulae \citep{Buemi_ska_chap}. In this context, the Wolf-Rayet star WR~75ab and the associated nebula PCG~11 provide an interesting case
study
in which the structure of a WR nebula and its interaction with the central star can be investigated in detail.

\pcg{} (PHR\,1633$-$4928) was discovered by \citet{Cohen2005} during the
follow-up of the Macquarie/AAO/Strasbourg H$\alpha$ (MASH) survey,
initially as a candidate planetary nebula. Optical spectroscopy
revealed a Population~I central star of type WN7h
\citep[\object{WR 75ab};][]{vdh06} at a distance of $4.1 \pm 0.4$~kpc,
surrounded by a nearly circular H$\alpha$ ring of radius
$36''$. Throughout this paper we adopt the Gaia-based
photogeometric distance posterior
of $3.9^{+0.4}_{-0.3}$~kpc \citep{BailerJones2021}, 
fully consistent with the estimate of
\citet{Cohen2005}. The optical nebula shows an unusual scalloped rim and a central region largely devoid of H$\alpha$ emission. The limited information available at other wavelengths, however, has so far prevented its three-dimensional structure and the physical origin of its scalloped rim from being established.
The only radio measurement available to
date was the detection at 843-MHz
from the Molonglo Observatory Synthesis Telescope (MOST), at a resolution
($43'' \times 56''$) insufficient to resolve the shell.  
In parallel with the present work, Filipovi\'c et al. (2026) investigate PCG 11 using ASKAP/EMU and survey MeerKAT observations, complemented by molecular-line data, with particular attention to its large scale radio and molecular environment.

In addition, the energetics and the
mass budget of PCG 11 in the context of our MeerKAT
survey of WR nebulae will be addressed in a companion paper
(Buemi et al., in prep.; hereafter Paper I); here we focus on
the new structural information: the resolved morphology,
the radio counterpart of the instabilities, and the three-dimensional geometry of the ionized nebula.

Here we present dedicated high-resolution radio continuum observations of PCG 11, obtained with MeerKAT at 2.625 and 3.062 GHz and with the Australia Telescope
Compact Array (ATCA) at 5.5 and 9.0 GHz, which allow us to resolve for the first time its detailed radio morphology, fragmentation, and interaction with the present-day stellar wind. 
The observations are described in Sect.~\ref{sec:obs} and the results
are presented in Sect.~\ref{sec:results}. In Sect.~\ref{sec:geometry} we model the
brightness distribution to constrain the geometry and orientation of the
nebula, and Sect.~\ref{sec:instab} we analyse the fragmented  structure
of the rim. In Sec.~\ref{sec:phys}, we investigate the physical conditions of the ionized gas and its interaction with the stellar wind. 
Our main conclusions are summarized in Sec.~\ref{sec:conclusions}.

%-------------------------------------------------------------------
\section{Observations and data reduction}
\label{sec:obs}

\subsection{MeerKAT S-band observations}
\pcg{} was observed with MeerKAT (project SCI-20230907-CB-02, 
 PI C.S. Buemi) 
on 2024 February 10 and 2024 April 12 in the S band, using the S2 and S4
sub-bands, centred at 2.6 and 3.1 ~GHz respectively, each with
875~MHz of bandwidth. 61 and 56 (S4, S2) antennas were used, for a total on-source
time of 22~min (Calibrators: PKS~1934-638, PMN~J1726-5529). The data were calibrated with the standard MeerKAT calibration pipeline (katsdpcal) and imaged by the MeerKAT Science Data Processor. 
The final images were corrected for the primary beam response. The
resulting synthesized beams and rms noise levels near the field
centre are listed in Table~\ref{tab:obs}.

We complement our observations with the 1.28-GHz image of the field
from the SMGPS , with an $8''$ circular beam and a local
rms noise of about 55~$\mu$Jy~beam$^{-1}$.

\subsection{ATCA observations}
\pcg{} was observed with the Australia Telescope Compact Array (ATCA)
in 2023 May (project C3526, PI C.S. Buemi) as part of a programme searching for
non-thermal radio emission from evolved high-mass circumstellar
environments. The observations were carried out in the 4\,cm band
with the CABB correlator in the CFB 1M (no zooms) mode, which
provides two 2-GHz wide IFs centred at 5.5 and 9.0~GHz (C and X
bands) observed simultaneously with 1-MHz channels. Two array
configurations were used, a 6\,km configuration and the compact EW352
configuration, for 5.25~h each per observing session over four
sessions. The combination of the two configurations provides
baselines from 31~m to 6~km, giving both sensitivity to extended
structure and subarcsecond to arcsecond scale resolution.
PKS-1934$-$638 was used as flux and bandpass calibrator and 1646-50
as phase calibrator. 
The data were reduced using MIRIAD \citep{Sault_95}, following standard procedures. The visibilities from the two configurations were combined in the $uv$-plane prior to imaging. Deconvolution was performed with the WSClean algorithm \citep{offringa-wsclean-2014,offringa-wsclean-2017}, utilizing Briggs intermediate robust weighting (robust = 0.5).
The final images have synthesized beams of
$2.92'' \times 1.55''$ (PA $= -12\degr$) at 5.5~GHz and
$1.86'' \times 0.93''$ (PA $= -8\degr$) at 9.0~GHz, with rms noise
levels of 38 and 27~$\mu$Jy/beam, respectively.

\begin{table}
\caption{Summary of the radio continuum observations of \pcg{}.}
\label{tab:obs}
\centering
\begin{tabular}{lccccc}
\hline\hline
Telescope & Band & $\nu_\mathrm{c}$ & Beam & PA & rms \\
          &      & (GHz) & ($\prime \prime $) & ($\degr$) & ($\mu$Jy/beam) \\
\hline
MeerKAT & L$^{(a)}$ & 1.3 & $8.0 \times 8.0$ & 0 &  55 \\
MeerKAT & S2 & 2.6 & $5.1 \times 4.7$ & 58 &  23 \\
MeerKAT & S4 & 3.1 & $4.4 \times 3.1$ & 60 & 14 \\
ATCA    & C  & 5.5   & $2.8 \times 1.2$ & $-11$ &  9 \\
ATCA    & X  & 9.0   & $1.9 \times 0.9$ & $-8$  & 8  \\
\hline
\end{tabular}
\tablefoot{The rms values are measured in source-free regions within
$\sim$2$'$ of the nebula. 
$^{(a)}$~Archival, from the SMGPS.
}
\end{table}

%-------------------------------------------------------------------
\section{Results}
\label{sec:results}

\subsection{Morphology}
\label{sec:morph}

\begin{figure*}
\includegraphics[scale=0.82]{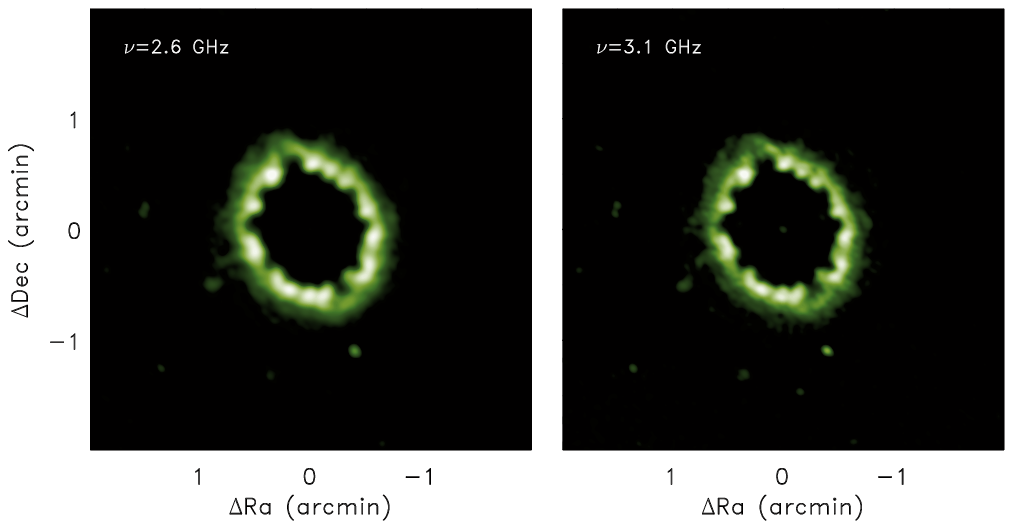}
\caption{MeerKAT S-band images of \pcg{} in the S2 (2.625~GHz, left)
and S4 (3.0625~GHz, right) sub-bands. The knotty structure of the rim,
with a characteristic separation matching the H$\alpha$
Rayleigh--Taylor scalloping of \citet{Cohen2005}, is clearly resolved.}
\label{fig:meerkat}
\end{figure*}

\begin{figure*}
\includegraphics[scale=0.82]{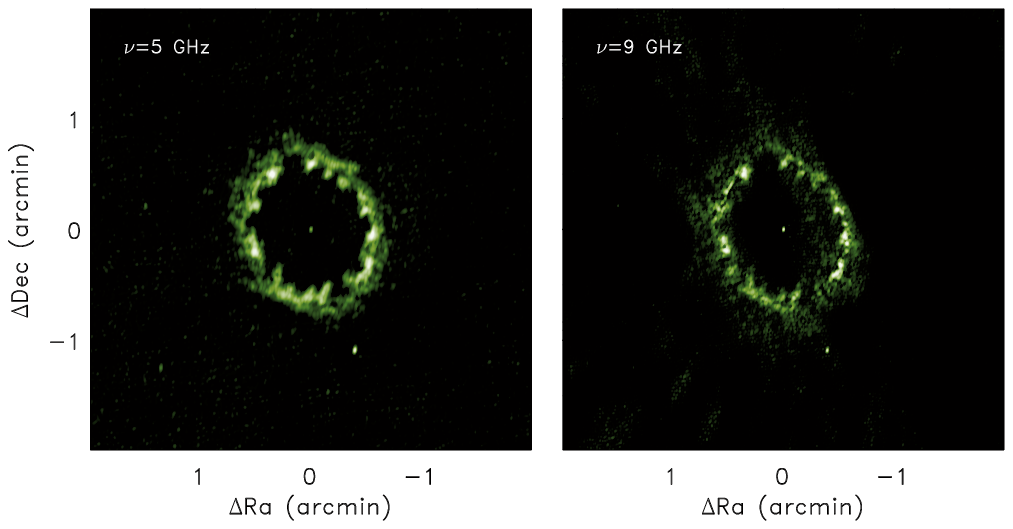}
\caption{ATCA images of \pcg{} in the C band (5.5~GHz, left) and X
band (9~GHz, right). The synthesized beam is shown in the bottom-left
corner. The central star is detected as an unresolved source in both
bands. Note the diffuse plume extending beyond the rim in the C-band
image.}
\label{fig:atca}
\end{figure*}

Figures~\ref{fig:meerkat} and \ref{fig:atca} show the MeerKAT and ATCA
images of \pcg{}. At all frequencies the nebula appears as a clumpy,
fragmented ring, in excellent agreement with the H$\alpha$ morphology
of \citet{Cohen2005}. The mean projected radius of the rim is $\simeq 36''$
(0.68~pc at 3.9~kpc) and its width, azimuthally averaged, is
$\simeq 11''$ in the MeerKAT images, only marginally affected
by the beam. The emission along the rim is concentrated in discrete
knots (Sect.~\ref{sec:instab}) separated by fainter emission; 
the presence of gaps in the radio ring is also apparent in the lower-resolution SMGPS image presented by Filipovi\'c et al. (2026).
The substantially higher angular resolution of the present observations, however, resolves the rim into individual condensations and finger-like structures
pointing towards the central star.

The interior of the ring shows no significant emission in any of the
images. Because the free-free emission is unaffected by dust
extinction, this deficit is a geometrical property of the nebula and
not an opacity effect; we quantify this point in
Sect.~\ref{sec:geometry}, where we show that it rules out a complete
spherical shell.

%___________________________________________________fig
%%%%%%%%%%%%%%%%%%%%%%%%%%%%%%%%%%%%%%%%%%%%%%%%%%%%%%%%%%%%%%%%%%%
%%%%%%%%%%%%%%%%%%%%%%%%%%%%%%%%%%%%%%%%%%%%%%%%%%%%%%%%%%%%%%%%%%%
%____________________________________________________________

%___________________________________________________fig
%%%%%%%%%%%%%%%%%%%%%%%%%%%%%%%%%%%%%%%%%%%%%%%%%%%%%%%%%%%%%%%%%%%
%%%%%%%%%%%%%%%%%%%%%%%%%%%%%%%%%%%%%%%%%%%%%%%%%%%%%%%%%%%%%%%%%%%
%____________________________________________________________

%___________________________________________________fig
%%%%%%%%%%%%%%%%%%%%%%%%%%%%%%%%%%%%%%%%%%%%%%%%%%%%%%%%%%%%%%%%%%%
\begin{figure}
\centering
\includegraphics[scale=0.18]{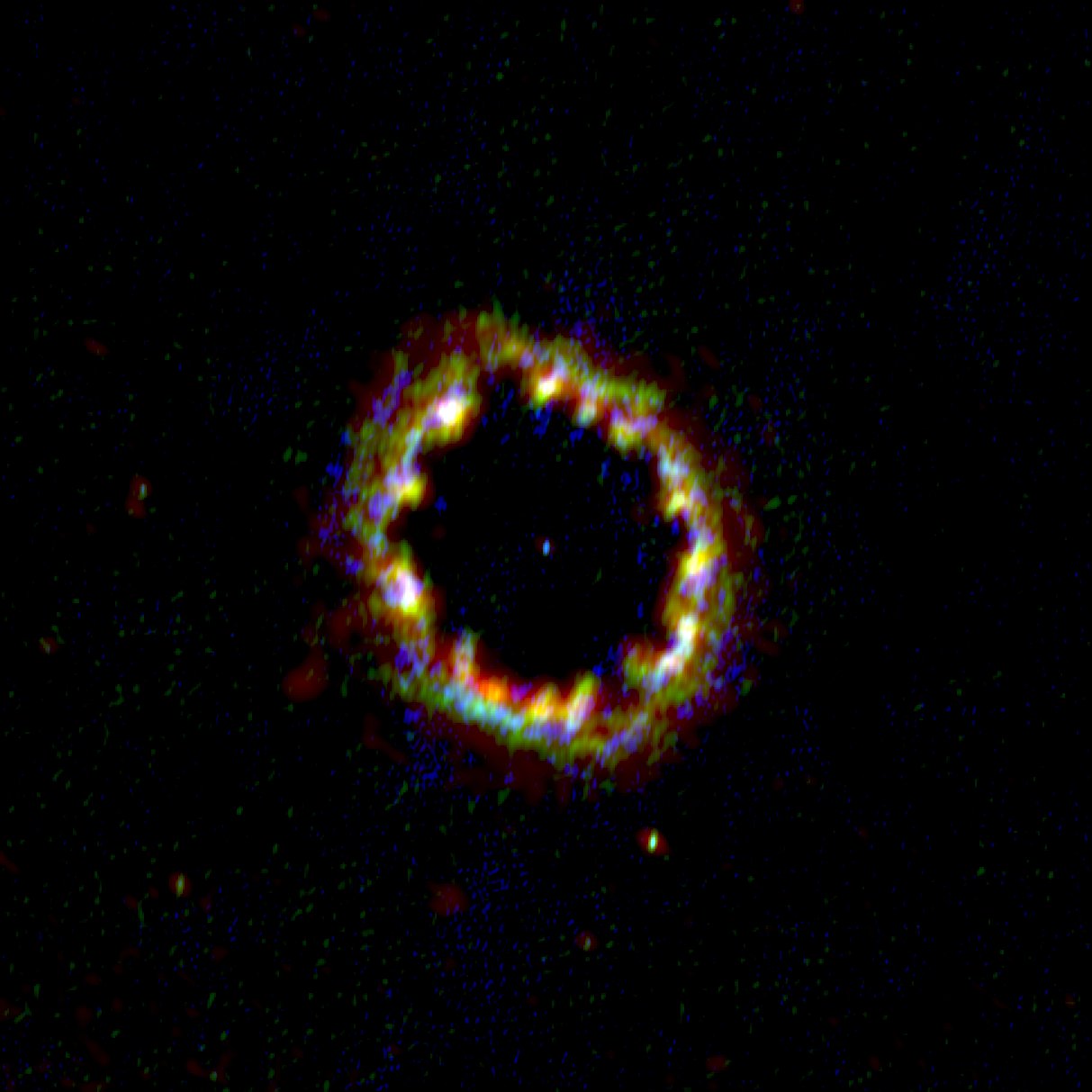}
\includegraphics[scale=0.18]{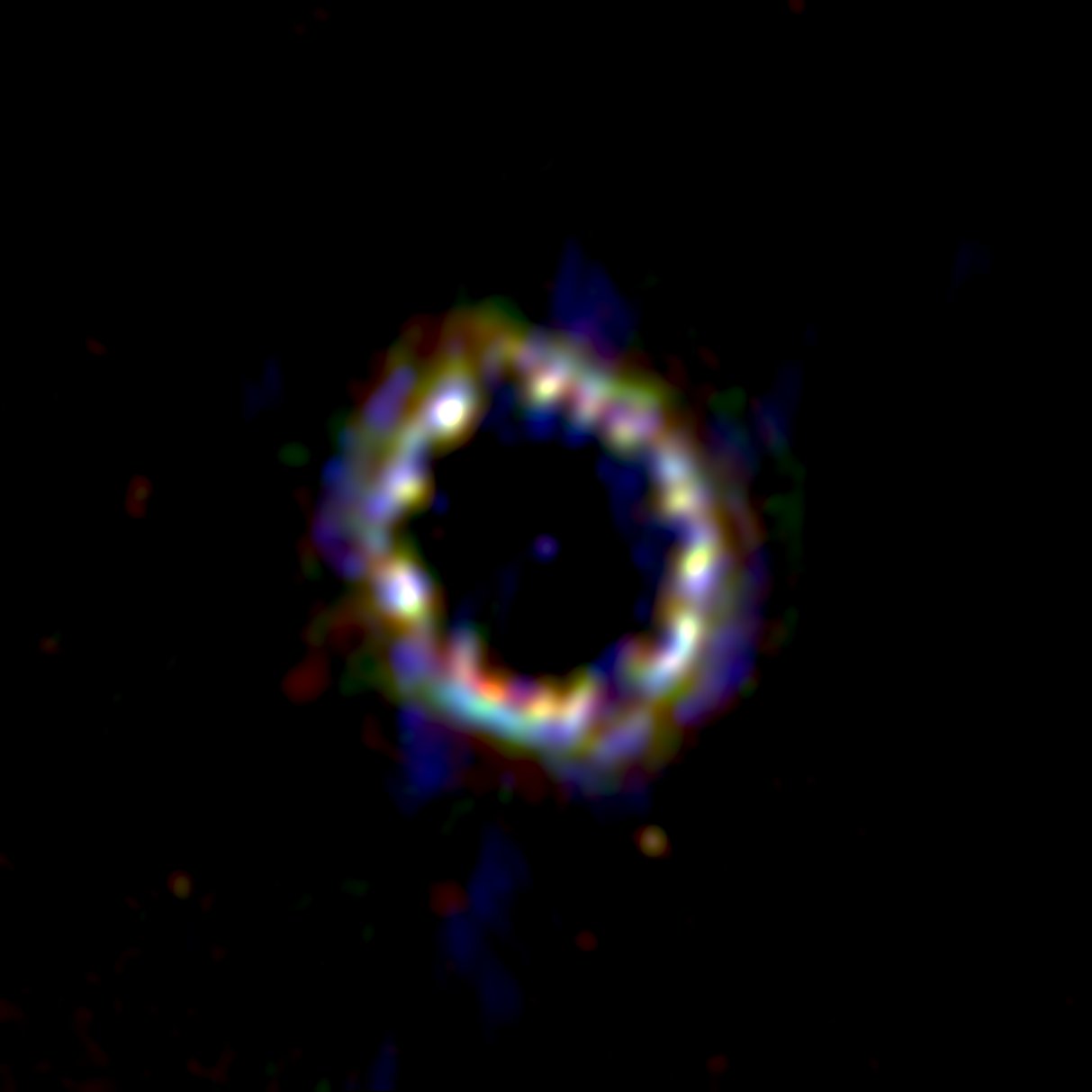}
\caption{
Three colour composite of the PCG 11 rim (red = S4, green = C, blue = X). Top panel, RGB image obtained using the native instrumental resolution. Bottom panel, RGB image obtained convolving the C and X maps with the  beam (larger) of the S4 map. 
}
\label{fig:rgb_fig}
\end{figure}
%___________________________________________________fig
%%%%%%%%%%%%%%%%%%%%%%%%%%%%%%%%%%%%%%%%%%%%%%%%%%%%%%%%%%%%%%%%%%%

The azimuthal brightness distribution of the rim is not uniform. Two
well-defined minima are present at PA $\simeq 20\degr$ and PA
$\simeq 85\degr$, coincident with the gaps seen in H$\alpha$
\citep{Cohen2005}; the southern rim, by contrast, is continuous and is
the brightest section of the ring. Beyond the rim, a
low-surface-brightness lobe ($\sim$13~$\mu$Jy~beam$^{-1}$ averaged
over the sector in the S4 image) extends to the south, spanning
PA $\simeq 150\degr$--$225\degr$ out to $r \ga 85''$, with no
counterpart on the north-eastern side. The same structure is
independently visible in the 1.28-GHz SMGPS image, obtained with a
different array configuration and imaging pipeline, which argues
against a deconvolution artefact. We return to the interpretation of
the rim knots and gaps in Sect.~\ref{sec:instab}.

\subsection{Integrated flux densities}
\label{sec:spix}

The integrated radio spectrum of WR 75ab was derived using the MeerKAT L-band continuum image together with the S-band sub-band  (S2 and S4) and the ATCA (C and X) images. To ensure a consistent comparison, the same source aperture was adopted at all frequencies. The outer radius r$_{\text{out}}$ of the nebula has been estimated as the distance where the average brightness of the radial profile at~1.3~GHz falls below the local noise level (RMS), 
estimated as the standard deviation of the pixel values within a nearby region free of significant emission.
The integrated flux density at each frequency was obtained 
through aperture photometry, using circular regions defined by r$_{\text{out}}$.
A local background correction was then applied to account for residual large-scale emission. 
The background level was estimated from a nearby source-free annulus located close to the nebula, and the mean surface brightness measured within this region was multiplied by the number of synthesized beams contained in the source aperture. This background contribution was subtracted from the integrated source flux.

The resulting background-subtracted integrated flux densities of the shell, and the flux densities (or upper limit) of the central star, are listed in Table~\ref{tab:fluxes}.
The uncertainties include the image rms and an absolute flux density calibration uncertainty of 5\%, added in quadrature.
The spectral index was defined according to $S_{\nu} \propto \nu^{\alpha}$, 
where $S_{\nu}$ is the radio surface brightness at frequency $\nu$. The  weighted least-squares fit performed using the three frequencies in logarithmic space, $\log S_{\nu} = \alpha \log \nu + C$, gives
$\alpha = 0.03 \pm 0.03$,
which indicates that the radio emission is consistent with optically thin thermal free-free emission. A predominantly thermal radio spectrum for PCG 11 was also found by Filipovi\'c et al. (2026), based on lower resolution radio continuum measurements.

\begin{table}
\caption{Integrated flux densities of the PCG\,11 nebula and of the central
star WR\,75ab.}
\label{tab:fluxes}
\centering
\begin{tabular}{ccc}
\hline\hline
$\nu_c$ (GHz) & $S_\nu$(ring) & $S_\nu$(WR\,75ab)$^{(a)}$ \\
\hline
1.3 & $66 \pm 3$~mJy & $\leq $ 105~$\mu$Jy \\
2.6 & $67 \pm 3$~mJy & $\leq $ 70~$\mu$Jy \\
3.1 & $67 \pm 3$~mJy & $89 \pm 22$~$\mu$Jy \\
5.5   & $70 \pm 4$~mJy & $ 110 \pm 21$~$\mu$Jy \\
9.0   & $69 \pm 4$~mJy & $ 212 \pm 20$~$\mu$Jy \\
\hline
\end{tabular}
\tablefoot{Uncertainties on the shell flux densities include the image rms
and a 5\% absolute flux density calibration uncertainty, added in quadrature.
\tablefoottext{a}{Point-source fit with FWHM fixed to the synthesized beam
and a free constant background.}}
\end{table}

\subsection{Detection of the central star}
\label{sec:star}

An unresolved source, spatially coincident with WR\,75ab in all
bands (fitted position $16^\mathrm{h}33^\mathrm{m}48.73^\mathrm{s}$,
$-49\degr28'44.3''$, within $0.3''$ of the optical position), is
detected in the S4, C and X images. The flux densities,  measured by
fitting a point source with the FWHM fixed to the synthesized beam and a free constant background,are reported in Table~\ref{tab:fluxes}
At 2.6 GHz, the source is not detected at a significant level. A weak positive feature is visible at the stellar position, but its significance is insufficient for a reliable flux density measurement. We therefore estimate the local rms directly within the ring and adopt a 3$\sigma$ upper limit of 75~$\mu\mathrm{Jy}$. Following the same proceedure, we adopt a 3$\sigma$ upper limit of 105~$\mu\mathrm{Jy}$ at 1.3~GHz, where no significant compact emission is detected at the stellar position.

This is the first radio detection of emission associated with the stellar wind of WR 75ab.
A power-law fit to the
three detections gives $\alpha_\star = 0.9 \pm 0.35$, consistent
within the uncertainties with the canonical $+0.6$ of a partially
optically thick $r^{-2}$ ionized wind \citep{Wright_Barlow1975};  spectral index steeper than the canonical value is commonly observed in WR winds (e.g. \citealp{Perry_26.549g1100P}) and may arise from radial variations in the wind properties, including a radially varying degree of clumping \citep{Nugis_98,Daley_Yates_16}.

Assuming that the compact radio emission is dominated by thermal free-free emission from the stellar wind, the mass-loss rate follows from
\citet{Wright_Barlow1975}:
\begin{equation}
\dot{M} = 0.095 \,
\frac{\mu \, v_\infty \, S_\nu^{3/4} \, d^{3/2}}
     {Z \, (\gamma \, g_\nu \, \nu)^{1/2}} \;\; \msunyr ,
\label{eq:wb}
\end{equation}
with $S_\nu$ in Jy, $d$ in kpc, $v_\infty$ in \kms{} and $\nu$ in Hz.
Adopting $v_\infty = 1350$~\kms{} \citep{Cohen2005}, $d = 3.9$~kpc, and
composition parameters for a WN7h star
($\mu = 2$, $Z = \gamma = 1$)
the 9~GHz flux density gives
$\dot{M}$ of $\sim 1.7 \times 10^{-5}\,\msunyr$. 
While Eq.~\ref{eq:wb} assumes a homogeneous wind, for a clumped wind with a volume filling factor of $f$, the inferred mass-loss rate needs to be reduced by $f^{1/2}$. 
Our adio-derived mass-loss rate is comparable to the value of $1.0 \pm 0.3 \times 10^{-5}\,\msunyr$ obtained by \citet{Cohen2005} from the dereddened 8.3~$\mu$m excess of WR 75ab. \cite{Crowther_07} reports a representative mass-loss rate of $\log(\dot M/M_\odot\,\mathrm{yr}^{-1})=-4.8$ for WN7 stars, based on stellar atmosphere analyses that account for wind clumping with a volume filling factor of $f=0.1$. Although this value is close to our estimate, the comparison is only indicative, since WR~75ab is classified as WN7h and the degree of clumping in the radio-emitting region of its wind remains unconstrained.
%We therefore conclude that our estimate of the mass-loss rate in WR~75ab is in agreement with that of $1.0 \pm 0.3 \times 10^{-5}\,\msunyr$ obtained by \citet{Cohen2005} from the dereddened 8.3~$\mu$m excess., 

%-------------------------------------------------------------------
\subsection{Radio spectral index map}
\label{sec:spectral_index}

\begin{figure*}
\centering
\includegraphics[width=0.99\textwidth]
{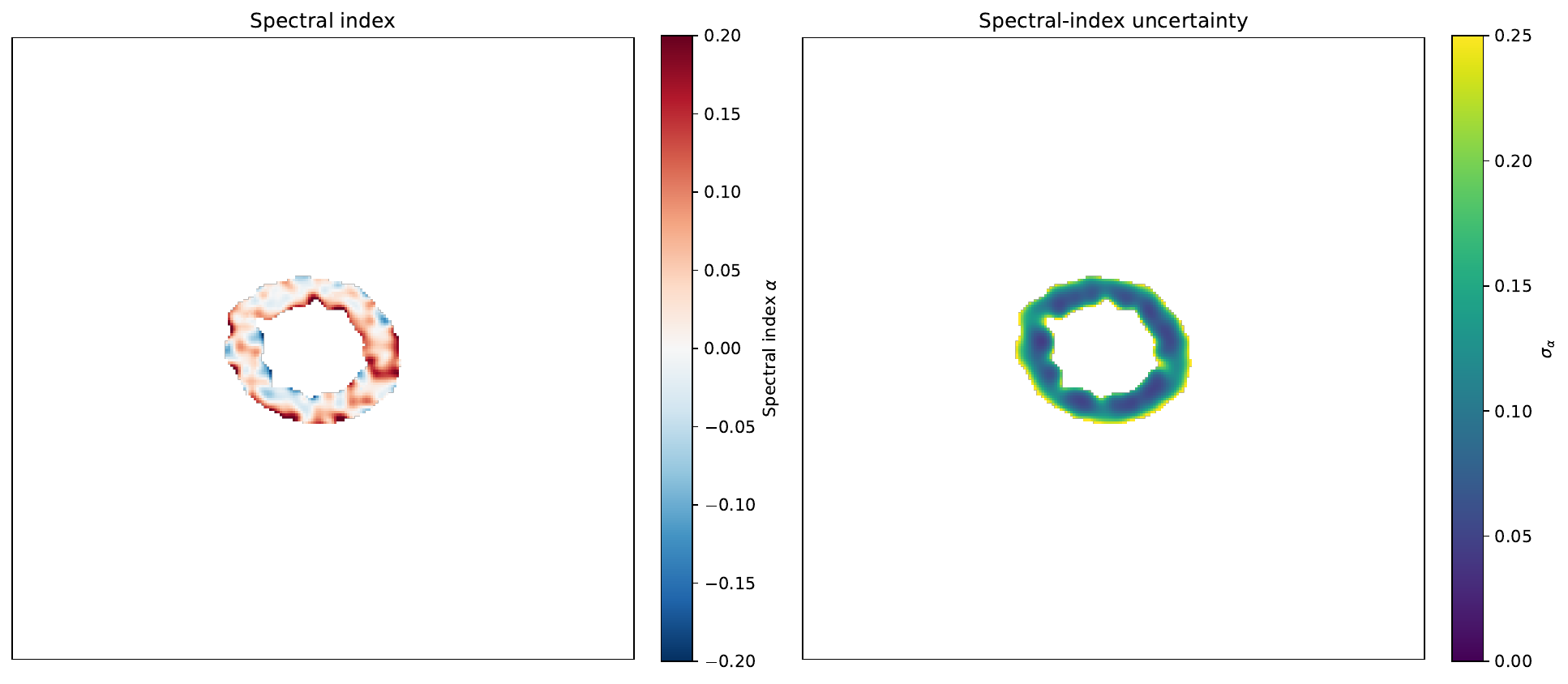}
\caption{Left: spectral-index map of PCG 11 derived from the L, S2 and S4 images. Right: 
corresponding spectral-index uncertainty map. The maps are in Galactic coordinates, the 
native grid of the SMGPS image to which the S2 and S4 images were reprojected (Sect.~\ref{sec:spectral_index}), 
and are therefore rotated with respect to the equatorial orientation of Figs.~\ref{fig:meerkat}--\ref{fig:rgb_fig}.}
\label{fig:alpha_ma}
\end{figure*}

\begin{figure}
\centering
\includegraphics[width=0.45\textwidth]{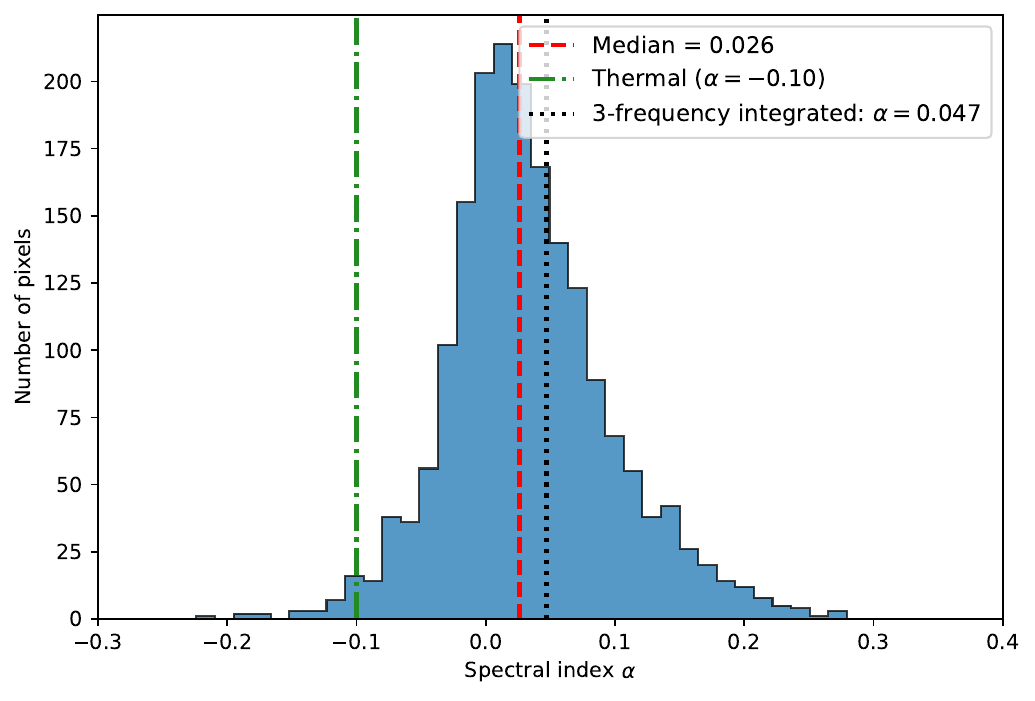}
\caption{
Distribution of pixel-by-pixel spectral index values $\alpha$ across 
the nebula, derived from the three frequency (L, S2, S4) fit. The red dashed 
line marks the median of the distribution; the black dotted line marks the 
spectral index obtained from the integrated flux densities at the three 
frequencies; the green dash-dotted line marks the value expected for 
optically thin thermal freefree emission. The distribution 
is unimodal and centred close to the thermal value, consistent with free-free 
emission from the nebula as a whole. 
}
\label{fig:alpha_hist}
\end{figure}

\begin{figure}
\centering
\includegraphics[width=0.45\textwidth]{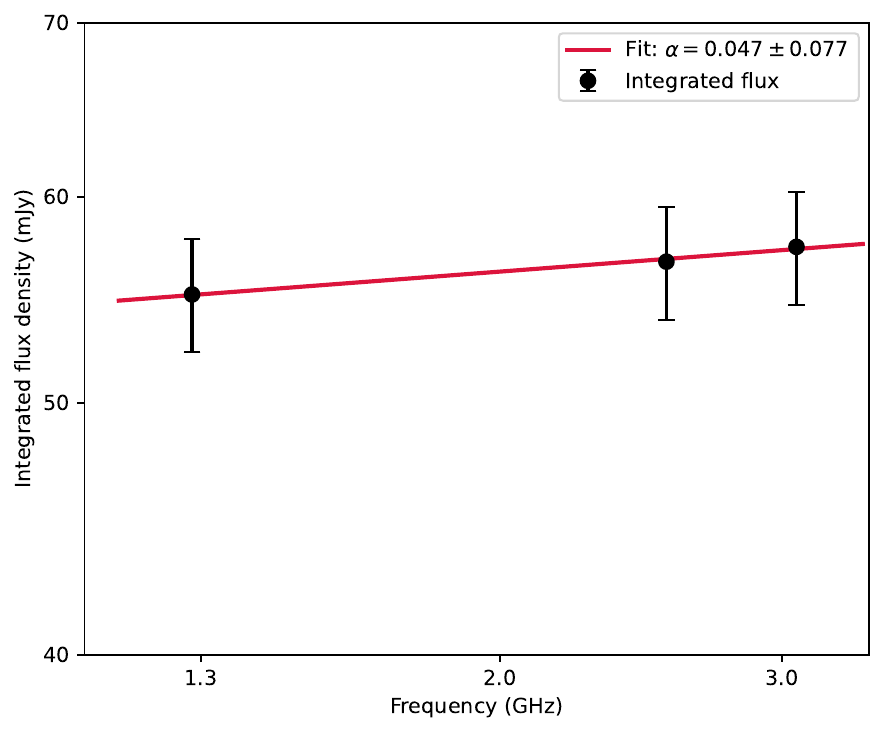}
\caption{Spectral index from integrated flux, derived from the integrated flux density measured only within the mask containing pixels above the 5$\sigma$ threshold in all three bands (see text). 
%\textit{Calcoli in 
%\texttt{WR75ab\_spectral\_index\_pulito.ipynb})}
}
\label{fig:integrated_spectrum}
\end{figure}

We investigated the spectral behaviour of the radio continuum emission associated with the WR~75ab nebula using the SMGPS at 1.284 GHz and our own MeerKat images at 2.625 and 3.062 GHz.
Since a reliable pixel-by-pixel comparison requires identical angular
resolution, astrometric sampling, and flux density units, the three images were
first brought to a common reference frame. The higher-resolution images were
convolved to a circular Gaussian beam with a full width at half maximum of
$8\arcsec$, corresponding to the lowest  
angular resolution available among
the datasets. 
%%%%%%
The convolution was followed by the appropriate beam-area correction, considering the ratio of beam area at different frequaencies, that is $\Omega_{L}/\Omega_{S2}=2.7$ and $\Omega_{L}/\Omega_{S4}=4.6$.
The images were then reprojected onto the same WCS and pixel grid
adopted for the 1.284~GHz map.

The relative astrometric alignment of the maps was examined both visually and
through image cross-correlation. The measured displacement was substantially
smaller than one pixel, with offsets of approximately $-0.07$ and $+0.13$ pixels
along the two image axes. Since these values are negligible compared with the
common $8\arcsec$ beam, no additional astrometric shift was applied. 

Only pixels with finite and positive surface brightness in all three images and
satisfying the adopted signal to noise threshold at every frequency were retained.  The final mask therefore contains only the extended shell emission associated with WR~75ab, excluding the compact source located south-west of the nebula by an appropriate mask.

The uncertainty on the spectral index, $\sigma_{\alpha}$, was estimated 
pixel-by-pixel by propagating the local flux density uncertainties through
the logarithmic fit. The uncertainty maps include the rms noise measured in the
individual radio images, while the resulting $\sigma_{\alpha}$ map was used to
identify regions in which apparent spectral variations may be dominated by low
signal-to-noise ratios. The median uncertainty over the nebular mask is
$\sigma_{\alpha,{\rm med}}\simeq 0.11$. Consequently, spectral-index variations
of only a few hundredths cannot be regarded as significant on an individual-pixel
basis.

The resulting three-frequency spectral-index map is shown in
Fig.~\ref{fig:alpha_ma}, together with the corresponding uncertainty map. The
emission over most of the shell is characterized by a nearly flat radio spectrum,
with the spectral-index distribution centred close to zero. The median value over
the nebula is $\alpha_{\rm med}$=$\,$0.026.

The map shows alternating regions with slightly positive and slightly negative
indices around the shell, but most of these fluctuations are comparable to the
local spectral index uncertainties. There is no extended region exhibiting the
strongly negative spectral index expected from a dominant synchrotron component.

The distribution of the valid pixel values is shown in
Fig.~\ref{fig:alpha_hist}. The histogram peaks close to
$\alpha\simeq 0.02$ and includes the reference value
$\alpha=-0.1$, expected for optically thin thermal free--free emission. A modest
asymmetric tail extends towards positive spectral indices, reaching values of
approximately $\alpha\simeq0.2$--$0.3$. Pixels belonging to this tail are not
uniformly distributed over the field, but occur in small, localized segments of
the shell, particularly close to some of its outer boundaries. Their spatial
coherence suggests that they do not arise exclusively from isolated random
outliers. However, they are limited in extent and are often located in regions
where the surface brightness is lower and the fitted spectral index is less
precisely constrained. We therefore do not interpret the positive tail as firm
evidence for a distinct physical emission component. It may instead result from a
combination of noise, residual background differences, small deconvolution
artefacts, and genuine but weak variations in the optical depth or density of the
ionized gas.

The median value $\alpha_{\rm med}$ is also consistent with the spectral index derived from the integrated flux density
measured within the mask common to the three maps, shown in Fig.~\ref{fig:integrated_spectrum}.
The agreement between the spatially resolved and
integrated measurements indicates that the integrated radio spectrum is not driven by
a small number of anomalous pixels or by the excluded compact source.

Interestingly, Fig.~\ref{fig:rgb_fig} suggests that some portions of the outer rim, particularly in the south-eastern and north-western sectors, appear relatively brighter in the C and X band images, even after convolution to a common beam. This may indicate a different spectral behaviour in these regions, possibly associated with locally optically thicker or more strongly compressed gas, shifting the free-free turnover to higher frequencies.

However, we do not extend the pixel-by-pixel spectral-index analysis to the ATCA bands: although the images can be convolved to a common restoring beam, differences in uv-coverage and spatial filtering between MeerKAT and ATCA, and between the ATCA C and X bands, prevent a sufficiently robust comparison of the highly structured shell on a pixel-by-pixel basis. These effects are much less relevant for flux densities integrated over the whole nebula.

Overall, the spectral-index maps, their associated uncertainty distribution, the
pixel histogram, and the integrated radio spectrum consistently indicate that the
radio continuum emission from the WR~75ab shell is predominantly thermal, consistent with the lower-resolution analysis of Filipovi\'c et al. (2026).
Our higher-resolution, three-frequency MeerKAT analysis further shows that this behaviour persists across the resolved shell and 
does not provide evidence for a significant non-thermal contribution. 

%-------------------------------------
\section{The geometry of the nebula}
\label{sec:geometry}
The radial brightness profiles presented in Paper~I already showed that a spherical shell cannot reproduce the observed central depression, whereas a ring-like geometry provides a significantly better description. Here we revisit this result using the higher angular resolution of the new MeerKAT and ATCA data, with the aim of constraining the three-dimensional geometry of the emitting shell rather than simply reproducing its radial profile. 
The radio morphology immediately evidences that the emission is concentrated in a narrow annular structure, suggesting that the ionized gas is confined to an equatorial layer. We therefore first estimate the inclination of this layer directly from its projected shape: for each position angle (PA) we locate the radius of peak surface brightness within a $\pm$5$^{\prime\prime}$ sector between 20$^{\prime\prime}$ and 60$^{\prime\prime}$, tracing the locus of the projected mid-plane of the shell, and fit the resulting (PA, r) points with the polar equation of an ellipse. Since the projection of a circular ring is an ellipse with axis ratio b/a = cos i, this yields the inclination directly: we obtain $a = 39.0^{\prime\prime}$, $b = 32.7^{\prime\prime}$, PA = 31.3$^{\circ}$, and $i = 33.0^{\circ}$, stable against the choice of initial parameters in the fit.

The azimuthally averaged radial profiles of the S4 band  were 
extracted about the central star emission in  1$^{\prime\prime}$ bins out to  75$^{\prime\prime}$, after subtracting a background level, derived as median intensity in a 70$^{\prime\prime}$ - 90$^{\prime\prime}$ annulus, and then normalized to its peak value. 

We then explored two simple axisymmetric geometries of a torus and a hollow cylinder and  then compared the results with the hollow shell geometry. Each geometry was projected assuming an optically thin, uniformly emissive medium at the inclination i = 33.0° determined above; for a given sky radius r and position angle $\phi$, the corresponding deprojected radius is $R(r,\phi) = r / \sqrt{\cos^2 \phi + \sin^2 \phi \cos^2i}$, and the model profile at r is obtained by averaging the emissivity over $\phi$. For the torus, we assumed the  profile 
$I(r) = 2 \sqrt{a^2 - (R(r,\phi) - R_0)^2}$
where $a$ is the tube radius of the torus, and $R_0$ is the distance from the star to the centre of the tube cross-section. 
For the hollow cylinder we assumed a top-hat profile of inner and outer radius $r_1$ and $r_2$ respectively. We also compared the fit with a hollow spherical shell of inner radius $R_1$ and outer radius $R_2$. Despite the radial profiles were extracted with a step size of $\Delta r = 1^{\prime \prime}$ to preserve spatial detail in the graphs, the goodness of fit ($\chi^2_r$) was evaluated considering the actual number of independent resolution elements ($N_{\text{eff}} \approx N_{\text{bins}} \times \Delta r / \text{FWHM}$).
	 
Each of these three profiles was convolved with a Gaussian representing the synthesized beam before being compared with the observed radial profile.
%-------------------------------------------------------------------
\begin{figure}
    \centering
    \includegraphics[width=\columnwidth]{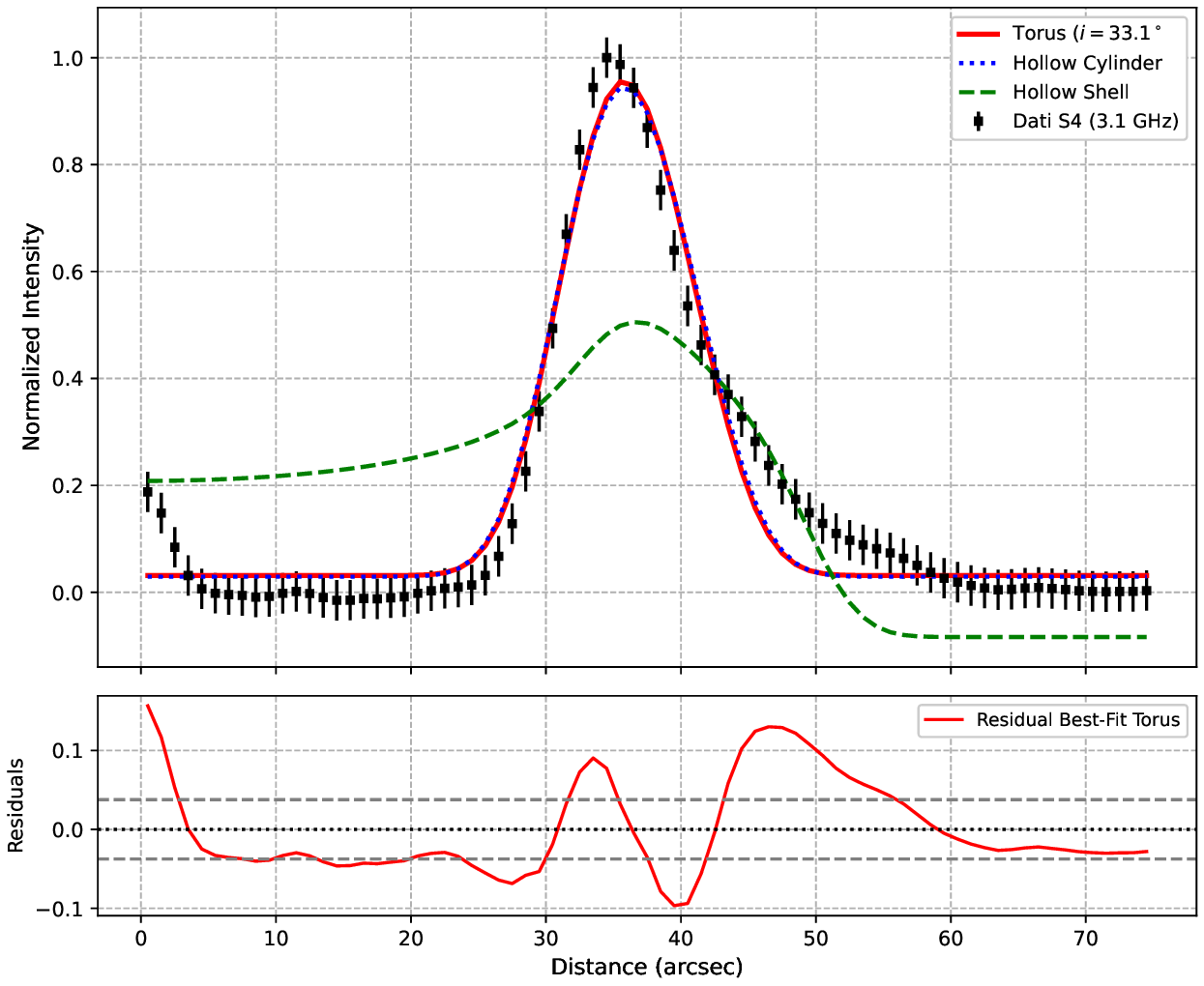}
    \caption{Azimuthally averaged radial brightness profile of the S4 (3.1 GHz) image (black points with 1$\sigma$ error bars), with the best-fitting models overlaid (top panel), and residuals of the best-fit torus model (bottom panel; grey dashed lines mark the $\pm1\sigma$ noise level).}
    \label{fig:radial_profile}
\end{figure}
%-------------------------------------------------------------------

%-------------------------------------------------------------------
\begin{table}
\caption{Geometric models fitted to the azimuthally averaged
S4 radial profile.}
\label{tab:geom}
\centering
\small
\setlength{\tabcolsep}{2.5pt}
\begin{tabular}{lcccc}
\hline\hline
Model & $i$ & \multicolumn{2}{c}{Best-fit parameters} & $\chi^2_r$ \\
\hline
Hollow sphere
 & -- 
 & $R_1=35.0\pm2.4''$
 & $R_2=47.9\pm2.4''$
 & 36.3 \\

Torus
 & $33.0^\circ$
 & $R_0=39.1\pm0.2''$
 & $a=6.7\pm0.5''$
 & 3.1 \\

Hollow cylinder
 & $33.0^\circ$
 & $r_1=33.7\pm0.4''$
 & $r_2=44.5\pm0.4''$
 & 3.1 \\
\hline
\end{tabular}

\tablefoot{$\chi^2_r$ was computed accounting for the correlation
between adjacent radial bins introduced by the synthesized beam.
Best-fit parameters were verified against multiple starting points to confirm convergence to the global $\chi^2$ minimum.}
\end{table}
%-------------------------------------------------------------------
%-------------------------------------------------------------------

The fit results are plotted in Fig.~\ref{fig:radial_profile} and the best-fitting parameters are listed in Table~\ref{tab:geom}. While the
spherical shell is strongly rejected, the torus  and the hollow cylinder
fit the profile equally well. 
We note that the morphology of the rim is also reminiscent of the equatorial ring of SN 1987A \citep{Burrows_95, Tegkelidis_24}, where the clumpy, knotted structure is likewise confined to the equatorial plane. This supports the interpretation of the scalloped rim of PCG 11 as belonging to an equatorial, torus-like structure, despite the different physical scales and origin of the ionization.

The present data constrain the geometry of the ionized shell but do not establish the origin of its equatorial enhancement. Possible explanations include anisotropic mass loss during an earlier evolutionary phase. For example, some bona fide Luminous Blue Variables, such as AG Car and HR Car, are known to be fast rotators, and this can produce an equatorially enhanced wind density structure \citep{Groh_11}: toroidal and bipolar structures are
often observed among LBV nebulae  in dust, ionized gas, and molecules (i.e., HR~Car studied by \citealp{Buemi2017}, and the equatorial ring in AG~Car  studied by \citealp{Bordiu2021}).
Other alterantive explanations that have to be considered include binary interaction or interaction with an anisotropic ambient medium. 
Distinguishing among these scenarios would require additional constraints on the stellar properties, nebular abundances, and gas kinematics. 

Irrespective of its origin, however, the geometric modelling shows that the observed ring-like morphology does not arise simply from the projection of a spherical shell, but reflects an intrinsically equatorially enhanced distribution of ionized material.

\section{The fragmented rim: radio counterpart of the instabilities}
\label{sec:instab}
%-------------------------------------------------------------------
\begin{figure}
    \centering
    \includegraphics[width=\columnwidth]{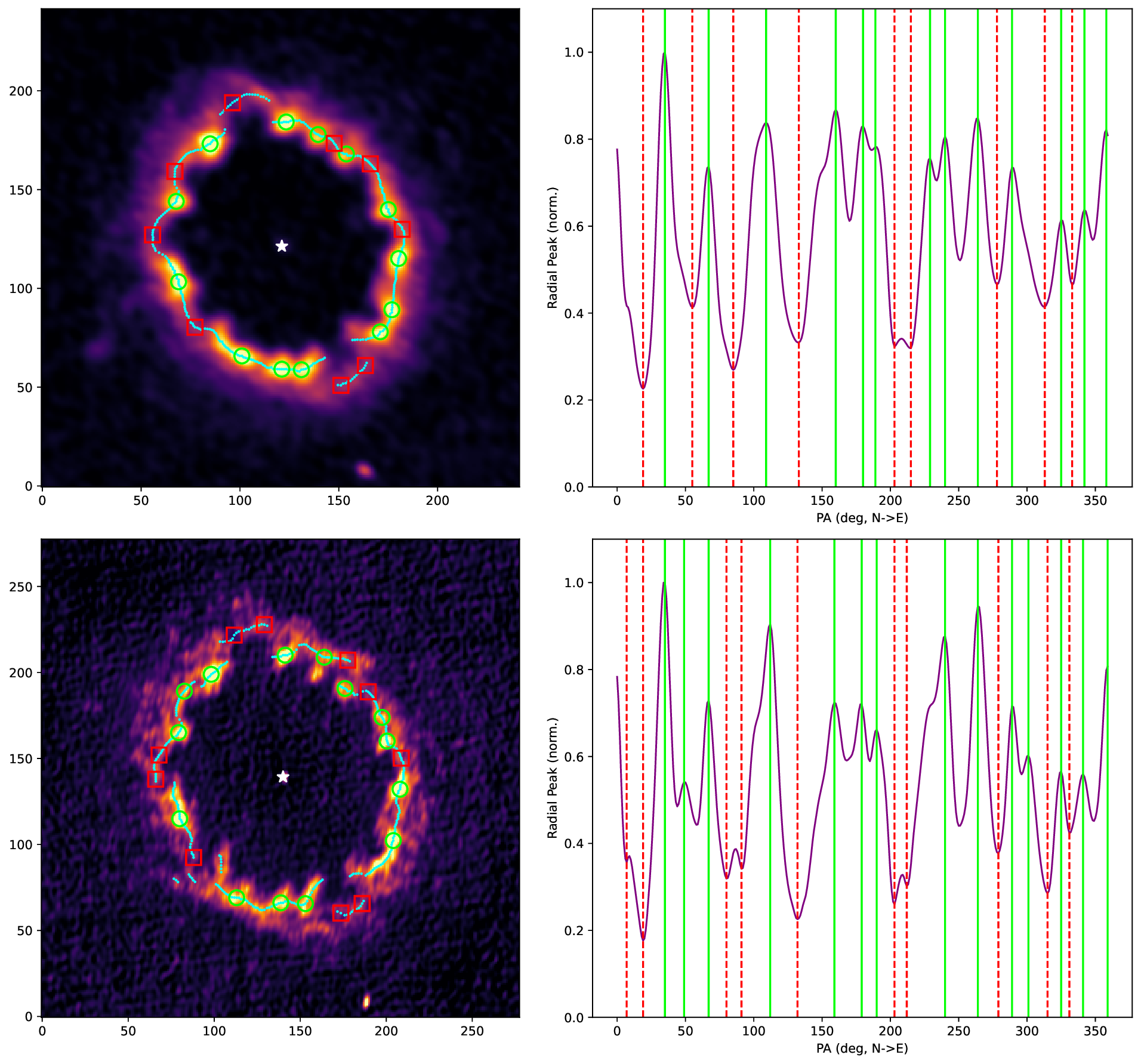}
    \caption{Azimuthal peak-tracing analysis of the ring rim in S4
(top) and C band (bottom). Left: surface-brightness map with the
traced rim (cyan dots) and the identified knots (green circles) and
gaps (red squares) overlaid. Right: corresponding azimuthal profile
of the radial peak brightness (not azimuthally averaged), with knot
and gap positions marked by solid green and dashed red lines,
respectively. 
}
    \label{fig:knots_gaps_profile}
\end{figure}
%-------------------------------------------------------------------
At 3.9~kpc the best-fitting torus tube diameter is
$13.4''\simeq0.25$~pc, notably the same scale as the 
Rayleigh-Taylor (RT) scalloping peak-to-peak distance % suggestion by Cristobal
($12''\pm1''$, 0.23~pc) measured in H$\alpha$
\citep{Cohen2005}.
The azimuthal brightness profile along the rim
(Fig.~\ref{fig:knots_gaps_profile}) is highly structured: we identify
$\simeq 14$ distinct knots (13 in S4, 14 in
C band), with a mean angular separation of
$\simeq 26^\circ$, i.e. $\simeq 16''$ (0.30 pc) along the ring.
Taking into account the smoothing of adjacent peaks by the $\sim$4$''$ beam, this is consistent with the $12'' \pm 1''$ characteristic wavelength of the H$\alpha$ scalloping.

At the resolution of the S4 and C band maps, a discrete population of individual condensations is resolved. These condensations are separated
by clear azimuthal gaps where the surface brightness drops. In the C band map, where the angular resolution is highest, these knots are not simply point-like brightness enhancements, but some of them visually appear elongated quite perpendicularly to the ring boundary. The high angular resolution of the radio observations thus provides a detailed, extinction-free view of the fragmentation of the ionized shell. In particular, the ATCA images trace the ionized gas on angular scales of only a few arcseconds, allowing to characterize the fragmentation in terms of its azimuthal distribution, characteristic spatial scale, and resolved morphology. These properties can be related to the independently constrained geometry and thickness of the shell, providing quantitative observational tests for future hydrodynamical modelling of wind-driven shell fragmentation. The morphology of elongated protrusions separated by low brightness gaps, strongly resembles the classic observational signature of Rayleigh-Taylor (RT) instability, in which perturbations at the interface between fluids of different densities grow into finger-like structures. \citet{Cohen2005} interpreted the regular H$\alpha$ scalloping and its finger-like morphology as the signature of RT instability. Together, the optical and radio observations provide a detailed multi-wavelength view of the fragmented structure of PCG~11.
The independent recovery of a comparable characteristic scale in the radio, together with the elongated morphology resolved by ATCA, strengthens this interpretation and shows that the regular pattern is an intrinsic property of the ionized shell. 

The quasi-regular distribution of the condensations further supports an instability-driven origin for the fragmentation, rather than a purely random distribution of pre-existing density inhomogeneities in the progenitor wind or in the ISM, as a purely clumpy, randomly fragmented shell is not expected to produce this systematic elongation. While RT instability provides a natural explanation for the large-scale fragmentation pattern, the observed morphology likely reflects the combined action of several physical processes operating during the  subsequent shell evolution

It is to note that the knots are distributed along the entire
circumference, as are the H$\alpha$ scallops, implying that the
instability operates coherently over the whole equatorial shell.
In addition, the characteristic spacing of the fragments is comparable to the ring thickness inferred indipendently from the geometric fits ($2a \simeq 0.25$ pc for the torus model), indicating that the shell thickness may set the characteristic length scale of the fragmentation. 
action of several physical processes operating during the  subsequent shell evolution.

%-------------------------------------------------------------------
%\section{Physical conditions of the ionized gas}
 \section{Interaction between stellar wind and fragmented shell}
\label{sec:phys}
The radio morphology and the equatorially enhanced distribution of the ionized gas described in the previous sections indicate that PCG~11 is no longer an intact swept-up shell, but a fragmented structure that remains exposed to the present-day stellar wind. We thus examine the interaction between the wind and the fragmented shell, through the local pressure balance,  the escape of ionizing photons and the dispersion of ionized material from the shell.

The emission measure, electron density, ionized mass and volume
filling factor of the shell are derived from the integrated radio
emission in Paper~I. Here we derive the local density of the
ionized skin from the
S4 and C bands surface brightness assuming optically thin free-free emission. 
The brightness temperature is thus estimated from  the observed flux density per beam as 
\begin{displaymath}
T_B = \frac{S_\nu c^2}{2 k \nu^2 \Omega_\mathrm{beam}}, 
\end{displaymath}
and the corresponding optical depht is $\tau_\nu = T_B/T_e$, related to the emission measure (EM) from the relation by \cite{Mezger_1967}:
\begin{equation}
\tau_\nu = 8.235\times10^{-2}\left(\frac{T_e}{\mathrm{K}}\right)^{-1.35}
\left(\frac{\nu}{\mathrm{GHz}}\right)^{-2.1}
\left(\frac{EM}{\mathrm{pc\,cm}^{-6}}\right).
\end{equation}
The electron density is then obtained from the line-of-sight
path length $L$ fixed by the torus geometry of Sect.~\ref{sec:geometry}
($L = 2a \simeq 0.25$~pc),
$
n_e = \sqrt{EM/L}.
$
Because the absolute density scale depends on the adopted background subtraction, electron temperature, and assumed line of sight depth, the derived values should be regarded as characteristic rather than precise 
local densities. Under these assumptions, and adopting $T_e = 10^4$~K, 
the 
ionized shell has a characteristic electron density of about 2$\times$10$^2\ {\rm cm^{-3}}$. Based on the azimuthal modulation of the radio surface brightness, the inferred electron density shows  azimuthal fluctuations of about a factor of 2, associated with the fragmented morphology. 
These values are consistent with the inner-edge density of the shell model
adopted in Paper~I ($n_\mathrm{e,0} \simeq 140$~cm$^{-3}$ with a
steep, $\alpha = 3.5$, radial density profile), once the different
volume definitions are taken into account.

The corresponding thermal pressure $P_\mathrm{th}/k = 2 n_\mathrm{e} T_\mathrm{e}$, is  $\simeq 4\times10^{6}$~K~cm$^{-3}$ for the estimated characteristic $n_e$, and is expected to locally vary across the shell within the same factor of approximately two. 
The detection of the stellar wind (Sect.~\ref{sec:star}) allows  to
 compare it with  the wind ram pressure at the
shell. For
$\dot{M} = 1.7\times10^{-5}\,\msunyr$ and
$v_\infty = 1350$~\kms{}, the momentum flux at $R_0 = 0.74$~pc is
\begin{displaymath}
P_\mathrm{w} = \frac{\dot{M} v_\infty}{4 \pi R_0^{2}}
\simeq 2.2\times10^{-9}\,\mathrm{dyn\,cm^{-2}}, 
\end{displaymath}
corresponding to $ P_\mathrm{w}/k \simeq 1.6\times10^{7}\,\mathrm{K\,cm^{-3}}$.
At the star-facing inner edge of the tube,
$R_0 - a \simeq 0.6$~pc, $P_\mathrm{w}$ it is a factor $\sim$1.5
higher.
The wind ram pressure therefore differ from the characteristic thermal pressure by a factor of $\sim$~4 (up to $\simeq$ 6 at the inner edge), 
suggesting that the present-day stellar wind is still able to influence the shaping of the shell. A clumped wind would reduce $P_{\rm w}$ by a factor $f^{1/2}$ (Sect.~\ref{sec:star}), so the values quoted here are upper limits.

The fragmented morphology could also affect the propagation of ionizing radiation.
The recombination rate implied by the integrated radio flux
corresponds to an absorbed ionizing photon rate of
$Q_\mathrm{abs} \simeq 8\times10^{46}$~s$^{-1}$, assuming an optically thin
free-free emission at 3~GHz and $T_\mathrm{e}=10^4$~K. The
average Lyman continuum output of WN7h stars is $Q_\mathrm{H} \sim 10^{49.4}$~s$^{-1}$\cite[]{Crowther_07}, so the shell processes only $\sim$0.3\% of the
stellar Lyman continuum. This indicates that PCG~11 is highly porous to ionizing radiation, with most ionizing photons escaping through the polar directions and through the gaps opened by the fragmented shell

When the shell fragments into discrete condensations, a
transfer of ionized material from the shell into the interior of
the nebula is expected. This transfer may result from a combination of photoionization-driven flows \citep[e.g.][]{Bertoldi1989,Bertoldi1990}, hydrodynamic ablation by the shocked stellar wind, and turbulent mixing at the shell-bubble interface \citep{Arthur_96,Pittard2001,dwarkadas23}. 
Our observations do not allow  to measure the associated mass-transfer rate, but the combination of a fragmented shell, pressure balance with the present WR wind and the large escape fraction of ionizing photons strongly suggests that that mass exchange between the shell and the nebular interior is ongoing.

On larger spatial scales, Filipovi\'c et al. (2026) identify an extended CO structure around PCG 11, which they interpret as tracing the cumulative effect of stellar-wind feedback over a much longer timescale. The compact ionized nebula investigated here instead traces a more recent mass-loss episode.

%-------------------------------------------------------------------
\section{Conclusions}
\label{sec:conclusions}

We have presented high resolution radio continuum images of the
ring nebula \pcg{} around the WN7h star WR\,75ab, obtained with
MeerKAT (2.625 and 3.062~GHz) and ATCA (5.5 and 9.0~GHz). Such dataset provides the highest-sensitivity and highest-dynamic-range extinction-free view of its ionized surrounding nebula to date, and resolve its fragmented structure at radio wavelengths down to the scale of its individual condensations and finger-like structures. The ATCA observations, in particular, resolve the fragmented rim on angular scales of a few arcseconds, allowing to resolve and quantify,  rather than simply identify morphologically, its characteristic structure.
The central star WR 75ab is detected for the first time at radio wavelengths. Its rising radio spectrum is consistent with thermal free-free emission from the stellar wind and yields a mass-loss rate of order 10$^{-5}\,M_\odot\,{\rm yr}^{-1}$, comparable to previous estimates obtained at infrared wavelengths.

The emerging picture from our analysis is that PCG~11 is not simply another 
Wolf-Rayet ring nebula. Rather than a classical spherical wind-blown bubble, it 
appears to be an intrinsically equatorial confined structure viewed close to its 
symmetry axis. The coincidence between the radio knots and the H$_{\alpha}$ 
scalloping shows that Rayleigh-Taylor instabilities dominate 
the nebular morphology. 

Viewed at a moderate
inclination ($i\simeq33^\circ$), closer to face-on than edge-on, the
equatorial structure is caught \emph{in the act} of destruption. Its fragmented, gap-riddled rim lets the star's ionizing
photons leak out at a rate roughly 50 times larger than its
open polar geometry alone would allow, suggesting that fragmentation has fundamentally altered the interaction between the central star and its surroundings.
The ionizing photon budget, together with the gap-riddled rim, indicates that the nebula has already become highly porous.
We therefore suggest that PCG 11 represents a late evolutionary stage of a wind-blown WR 
circumstellar structure in which fragmentation and ionizing photons have become dynamically important.

Irrespective of the origin of its equatorial geometry, PCG~11 is therefore  a rare snapshot of a WR nebula in the process of distruption under the radiative and mechanical feedback of its own central star. 
Radio observations of evolved WR nebulae are still sufficiently rare that this transitional phase has remained poorly characterized observationally. 
Beyond the specific case of PCG~11, these observations provide insight into the late evolution of wind-blown bubbles around massive stars, as they could provide useful constraints  on models describing the development and subsequent evolution of fragmentation in circumstellar shells

Owing to its nearly pole-on orientation and the extinction-free radio continuum,  our target
provides an unusually clear view of processes that are probably common, but much more difficult to recognize, in other WR nebulae.

\begin{acknowledgements}
The MeerKAT telescope is operated by the South African Radio Astronomy Observatory, which is a facility of the National Research Foundation, an agency of the Department of Science, Technology and Innovation. This work has made use of the “MPIfR S-band receiver system” designed, constructed and maintained by funding of the MPI für Radioastronomy and the Max-Planck-Society.

The Australia Telescope Compact Array is part of the Australia Telescope National Facility (https://ror.org/05qajvd42), which is funded by the Australian Government for operation as a National Facility managed by CSIRO. We acknowledge the Gomeroi people as the Traditional Owners of the Observatory site.

CB acknowledges financial support from grant CEX2021-001131-S funded by MICIU/AEI/10.13039/501100011033 and from grant INFRA24023 (CSIC4SKA) funded by CSIC

CSB used ChatGPT (OpenAI) for language and editorial assistance during the preparation of this manuscript, including improvements to clarity, readability, and text structure. All scientific content, analysis, interpretation and conclusions are the sole responsibility of the authors.
\end{acknowledgements}

%___________________________________________________ bibliography
%%%%%%%%%%%%%%%%%%%%%%%%%%%%%%%%%%%%%%%%%%%%%%%%%xs
\bibliographystyle{aa} % style aa.bst
\bibliography{biblio}
%%%%%%%%%%%%%%%%%%%%%%%%%%%%%%%%%%%%%%%%%%%%%%%%%xs
%_________________________________________________END  bibliography
\end{document}